\pdfoutput=1
\documentclass[conference]{IEEEtran}
\IEEEoverridecommandlockouts
\PassOptionsToPackage{table, x11names, dvipsnames, svgnames}{xcolor}
\usepackage{cite}
\usepackage{bbm}
\usepackage{amsmath,amssymb,amsfonts}
\usepackage{algorithmic}
\usepackage[ruled,vlined,linesnumbered]{algorithm2e}
\usepackage{graphicx}
\usepackage{datatool}
\usepackage{textcomp}
\usepackage{csquotes}
\usepackage{xcolor}
\usepackage{xspace}
\usepackage[english]{babel}
\usepackage[acronym,nonumberlist]{glossaries}
\usepackage{glossaries-prefix}
\usepackage{amssymb}
\usepackage{standalone}
\usepackage{bm}
\usepackage{siunitx}
\usepackage{multirow}
\usepackage{tabularray}
\usepackage{booktabs, tabularx}
\usepackage[]{yquant}
\usetikzlibrary{quotes, fit}
\usepackage{braket}
\usepackage{textcomp}

\usepackage{hyperref}
\hypersetup{%
linktocpage=true, 
colorlinks=false,
pdfborder={0 0 0}, 
breaklinks=true, pdfpagemode=UseNone, pdfpagemode=UseOutlines,%
plainpages=false, bookmarksnumbered, bookmarksopen=true, bookmarksopenlevel=1,%
hypertexnames=true, pdfhighlight=/O,
pdftitle={Guided-SPSA},%
pdfauthor={Maniraman Periyasam},%
pdfsubject={QML},%
pdfkeywords={},%
pdfcreator={pdfLaTeX},%
pdfproducer={LaTeX with hyperref}%
}

\usepackage{mathtools}

\DeclarePairedDelimiter\floor{\lfloor}{\rfloor}

\newcommand{\ie}{\emph{i.e.}\xspace}

\newcommand{\etal}{\emph{et al.}\xspace}

\usepackage[inline]{enumitem}

\setlist*[enumerate]{label=(\roman*)} 

\usepackage[capitalise]{cleveref}

\newacronym[shortplural=GMMs]{GMM}{GMM}{Gaussian mixture model}
\newacronym[shortplural=HMMs]{HMM}{HMM}{hidden Markov model}
\newacronym[shortplural=DNNs]{DNN}{DNN}{deep neural network}
\newacronym[shortplural=SVDs]{SVD}{SVD}{singular value decomposition}
\newacronym[
    prefixfirst={a\ },
    prefix={an\ }
]{MCTS}{MCTS}{Monte Carlo tree search}
\newacronym[prefixfirst={a\ },prefix={an\ }]{MDP}{MDP}{Markov decision process}
\newacronym{CMDP}{CMDP}{constrained Markov decision process}
\newacronym{RL}{RL}{reinforcement learning}
\newacronym[shortplural=DTs]{DT}{DT}{decision tree}
\newacronym{SMT}{SMT}{satisfiability modulo theories}
\newacronym{IL}{IL}{Imitation Learning}
\newacronym[shortplural=CNNs]{CNN}{CNN}{convolutional neural network}
\newacronym[shortplural=DQNs]{DQN}{DQN}{deep Q-network}
\newacronym{AI}{AI}{artificial intelligence}
\newacronym{PPO}{PPO}{proximal policy optimization}
\newacronym{ML}{ML}{machine learning}
\newacronym{QML}{QML}{quantum machine learning}
\newacronym{NISQ}{NISQ}{noisy intermediate scale quantum}
\newacronym{QC}{QC}{quantum circuit}
\newacronym{VQC}{VQC}{variational quantum circuit}
\newacronym{VQA}{VQA}{variational quantum algorithm}
\newacronym{MNIST}{MNIST}{modified national institute of standards and technology}
\newacronym{FIM}{FIM}{Fisher information matrix}
\newacronym{IDU}{IDU}{incremental data-uploading}
\newacronym{DRU}{DRU}{data re-uploading}
\newacronym{QRL}{QRL}{quantum reinforcement learning}
\newacronym{SPSA}{SPSA}{simultaneous perturbation stochastic approximation}
\newacronym{SGD}{SGD}{stochastic gradient descent}
\newacronym{QEM}{QEM}{quantum error mitigation}
\newacronym{QPU}{QPU}{quantum processing unit}

\begin{document}



\title{Guided-SPSA: Simultaneous Perturbation Stochastic Approximation assisted by the Parameter Shift Rule
\thanks{

This work was supported by the German Federal Ministry of Education and Research (BMBF), funding program “quantum technologies from basic research to market”, via the project QLindA, grant numbers 13N15645 and 13N15647. WM acknowledges support by the High-Tech Agenda of the Free State of Bavaria.\\
email address for correspondence: 
maniraman.periyasamy@iis.fraunhofer.de}
}

\author{
    \IEEEauthorblockN{Maniraman Periyasamy\IEEEauthorrefmark{1}\IEEEauthorrefmark{2}, Axel Plinge\IEEEauthorrefmark{1}, Christopher Mutschler\IEEEauthorrefmark{1}, Daniel D.\ Scherer\IEEEauthorrefmark{1}, Wolfgang Mauerer\IEEEauthorrefmark{2}\IEEEauthorrefmark{3}
    }
    \IEEEauthorblockA{
        \IEEEauthorrefmark{1}Fraunhofer IIS, Fraunhofer Institute for Integrated Circuits IIS, Nürnberg, Germany\\
        \IEEEauthorrefmark{2}Technical University of Applied Sciences Regensburg, Regensburg, Germany\\
        \IEEEauthorrefmark{3}Siemens AG, Corporate Research, Munich, Germany
    }
}

\maketitle

\begin{abstract}
The study of variational quantum algorithms (VQCs) has received significant attention from the quantum computing community in recent years. These hybrid algorithms, utilizing both classical and quantum components, are well-suited for noisy intermediate-scale quantum devices. Though estimating exact gradients using the parameter-shift rule to optimize the VQCs is realizable in NISQ devices, they do not scale well for larger problem sizes. The computational complexity, in terms of the number of circuit evaluations required for gradient estimation by the parameter-shift rule, scales linearly with the number of parameters in VQCs. On the other hand, techniques that approximate the gradients of the VQCs, such as the simultaneous perturbation stochastic approximation (SPSA), do not scale with the number of parameters but struggle with instability and often attain suboptimal solutions. In this work, we introduce a novel gradient estimation approach called Guided-SPSA, which meaningfully combines the parameter-shift rule and SPSA-based gradient approximation. The Guided-SPSA results in a 15\% to 25\% reduction in the number of circuit evaluations required during training for a similar or better optimality of the solution found compared to the parameter-shift rule. The Guided-SPSA outperforms standard SPSA in all scenarios and outperforms the parameter-shift rule in scenarios such as suboptimal initialization of the parameters. We demonstrate numerically the performance of Guided-SPSA on different paradigms of quantum machine learning, such as regression, classification, and reinforcement learning.
\end{abstract}


\glsresetall
\section{Introduction}
\label{sec:introduction}

Machine learning has witnessed a surge of interest in numerous fields~\cite{Voulodimos2018,Otter2017,Vaswani2017,Baheri2020,fatima2017survey,winker:23:sigmodtutorial,bayerstadler:21:}. However, 
the underlying algorithms usually demand substantial computational resources and energy, which is troublesome for steadily increasing amounts of data. Quantum computing emerges as a potential alternative based on a fundamentally different computational paradigm~\cite{Wiedmann2023}. By harnessing the unique properties of quantum mechanics, fault-tolerant quantum computers are known to exhibit substantial speedups over classical computers for specialized problems~\cite{Shor94,Harrow09,pirnay2024superpoly}.
Such advantages are also known for machine learning tasks~\cite{Rebentrost:2014,Tychola2023}, and are augmented
by the promise of generalization based on 
fewer data points than classical methods~\cite{Caro:2022}, or improved accuracy~\cite{Liu_2021} (albeit achieving concrete
practical advantages over established classical heuristics
is a subject of intense debate~\cite{bowles2024better})

Current noisy intermediate-scale quantum (NISQ) devices in their earlier stages of development suffer from limited qubit counts, short coherence times, and high gate- and readout-error rates~\cite{preskill2018quantum,wintersperger:23:neutralatom}. This imposes restrictions on the size and complexity of quantum circuits, limiting algorithmic design choices.
%
Another challenge for \gls{QML} is the volume of data encountered in machine learning tasks:
Some approaches like quantum kernel methods \cite{schuld2021supervised} scale quadratically with the size of the dataset, which makes many methods impractical for large datasets.

A widely employed approach to overcome these challenges involves hybrid quantum-classical algorithms that combine classical and quantum computing. One prominent example are \glspl{VQA}~\cite{cerezo2021variational}. These utilize \glspl{VQC}, where specific parameters within some of the gates of the circuit can be adjusted. A classical optimization routine is used to tune the set of parameters of the \gls{VQC} to minimize the predefined objective function. This process entails \enquote{training} the \gls{VQC} by optimizing free parameters, which is a crucial ingredient for their
capabilities.

Gradient-based and gradient-free methods are two main approaches used for optimizing the \gls{VQC}s in \gls{VQA}s. Gradient-based methods, such as the parameter-shift rule, offer fast and stable convergence but scale linearly with the number of circuit parameters. The computational complexity in terms of the number of circuits evaluated (hereinafter referred to as computation complexity) for the parameter-shift rule~\cite{crooks2019gradients} is given by $O(2NM)$ where $N$ is the number of training data, and $M$ is the number of parameters in the circuit. Conversely, gradient-free methods do not necessarily demand linear scaling with the number of parameters but are less efficient in terms of convergence speed and accuracy. In order to overcome these shortcomings, the research question we pose in this study is as follows: Can we develop a gradient estimation technique for \gls{VQA}s that satisfies the following conditions:
\begin{enumerate*}
    \item it achieves stable convergence with a significantly lower number of circuit evaluations compared to existing methods,
    \item is realizable on current quantum hardware,
    \item is suitable for \gls{NISQ} devices (\ie, does not increase the circuit complexity in terms of gate count or the number of measurements required compared to other gradient estimation techniques like the parameter-shift rule).
\end{enumerate*}



To answer our research question, we introduce and explore a novel, efficient gradient estimation technique named Guided-SPSA that leverages the advantages of both parameter-shift rule and \gls{SPSA} by meaningfully combining them during the training process. The Guided-SPSA is realizable on quantum hardware. We evaluate our approach on different paradigms of \gls{QML}, such as regression, classification, and reinforcement learning, along with different problem setups such as noisy simulation, and suboptimal initialization

The rest of this paper is structured as follows: 
\cref{sec:related_work} provides an overview of related literature on different gradient estimation techniques for \gls{VQC}s. \cref{sec:Theory} introduces the relevant theoretical background, providing a formal introduction to \gls{VQC}s, encoding techniques, parameter-shift rule, and \gls{SPSA}. \cref{sec:guided-spsa} introduces the Guided-SPSA technique. \cref{sec:Results} gives a detailed description of various numerical experiments conducted to enunciate the performance of Guided-SPSA under different \gls{QML} paradigms and their results.

\section{Related Work}
\label{sec:related_work}

\gls{VQC}s are promising for solving machine learning and optimization problems using near-term quantum devices. Efficiently estimating gradients for these circuits is crucial for finding optimal parameters. Various techniques have been proposed in the literature to tackle this challenge of gradient estimation. One widely used method is the parameter-shift rule \cite{crooks2019gradients, Schuld19}, which estimates gradients by shifting the circuit parameters and measuring the changes in expectation values with respect to an observable. Despite being popular, the parameter-shift rule is computationally expensive, particularly for circuits with many parameters. Another common technique to compute gradients is through finite differences, which involves evaluating circuits at slightly different parameter values. Though certain variations of this method, such as the forward distance method, are not computationally expensive like the parameter-shift rule, they are vulnerable to numerical errors, especially in noisy environments. Linear combination of unitary gradients presented by Schuld~\etal~\cite{Schuld19} is another method used to estimate the gradients of a given VQC. Here, the gradient is estimated via linearly combining values obtained by evaluating the observable using parameter values perturbed about their forward-pass values. While the linear combination gradient estimator offers computational efficiency compared to the parameter-shift rule, it also suffers from numerical errors and inaccuracies. The adjoint method, introduced by Luo~\etal \cite{Luo_2020}, calculates gradients by applying the adjoint of the circuit to the measurement operator, eliminating the need for finite differences or multiple circuit evaluations. Meyer~\etal \cite{meyer2024qiskittorchmodule} reformulated this adjoint method to allow batch-parallelization. The efficiency of the adjoint methods generally depends on the ability to compute the adjoint of the circuit accurately, and cannot directly be executed on quantum hardware.

In recent years, \gls{SPSA}, introduced by Spall \etal~\cite{Spall92} have gained traction for their efficiency and robustness in noisy environments. Wiedmann~\etal  \cite{Wiedmann2023} studied the best-suited classical optimizers for parameter updates when using the SPSA-based gradient estimation technique. SPSA is 
an estimation method that approximates gradients by iteratively shifting all the parameters simultaneously with random shift values. It requires a large number of measurements for accurate gradient approximation. More sophisticated methods for optimizing \gls{VQC}s beyond the above methods have been introduced: One notable approach leverages Bayesian inference to efficiently estimate gradients presented by Bittel~\etal~\cite{bittel2022fast}.

Quantum Natural Gradient methods have recently gained attention, as they can efficiently navigate the parameter space by incorporating the geometry of quantum state manifolds~\cite{Stokes_2020}. These methods often show better convergence properties than more straightforward alternatives. However, the accurate computation and storage of the second-order information can be computationally and memory-intensive, especially for large-scale circuits. Another method that suffers from a similar bottleneck is trust region optimization \cite{Yuan1999}. Here, at every step, a sup-optimization routine is computed with the trust region, which involves Hessian estimation. Beyond gradient estimation techniques for NISQ devices, novel approaches such as quantum gradient evaluation through quantum non-demolition measurements suitable for error-corrected post-NISQ devices are also being explored~\cite{Solinas_2023}. Gradient-free optimization techniques are becoming increasingly popular as alternative methods for optimizing \gls{VQC}s without relying on gradient information. These techniques use heuristic search algorithms like evolutionary algorithms \cite{Chen_2022}, simulated annealing, or the use of gradient-free optimizers such as Nelder-Mead, Powell, and COBYLA~\cite{ostaszewski2021reinforcement} algorithms to explore the parameter space. Although gradient-free methods avoid the computational burden of gradient computation, they may require a significant number of circuit evaluations to converge to optimal solutions, especially for high-dimensional parameter spaces or noisy quantum devices. Additionally, these methods may struggle to escape local minima and cannot guarantee convergence to the global optimum.

Apart from the optimization techniques, improvements in quantum hardware capabilities such as increasing the number of available qubits, improving coherence times, reducing error rates, or adapting \gls{QPU} designs~\cite{safi:23:codesign,wintersperger:22:codes}, 
and efficient use of existing NISQ hardware~\cite{krueger:20:icse,sax:20:approximate,schoenberger:2023} are necessary to enable QML (abnd other quantum methods) to tackle more complex problems and handle larger datasets. Advanced training algorithms, such as Distributed Coordinate Descent \cite{Koyasu2023}, handle large datasets by letting multiple NISQ devices work on subsets of data concurrently and use the NISQ devices more efficiently.

Thus estimating gradients for \gls{VQC}s remains a topic of ongoing research, with different techniques offering various benefits and drawbacks in terms of computational efficiency, accuracy, and applicability to different circuit structures and noise models~\cite{greiwe:23:imperfections}. The selection of a particular method depends on circuit complexity, noise characteristics, and available computational resources. Further improvements in this field are crucial for unleashing the full potential of \gls{VQA} on near-term quantum devices, and also for gaining a better understanding of the source of computational power in quantum-classical algorithms.
\section{Theoretical Background}
\label{sec:Theory}
As we aim at a synthesis of \gls{SPSA} and parameter shift techniques to improve the performance of \gls{VQC}s, it seems appropriate to briefly summarize the theoretical background for each of these ingredients. Furthermore, we give a brief example of a gradient descent algorithm.

\subsection{Variational Quantum Circuits}
\label{sec:vqc}
\Gls{VQC}s are quantum circuits constructed using a mix of parameterized and non-parameterized unitary gates with tunable parameters. The parameters of the \gls{VQC}s are tuned to approximate a target function via some training algorithm. These circuits offer a flexible framework for solving various machine learning and optimization tasks.

\subsubsection{Structure of a VQC}

\begin{figure}[!bt]
    \centering
    \resizebox{0.9\linewidth}{!}{
    \begin{tikzpicture}
        \begin{yquant}
            qubit {$\ket0$} q[4];
            hspace {5mm} -;
            [name=e-top]
            box {$R_x(x_{0})$} q[0];
            box {$R_x(x_{1})$} q[1];
            box {$R_x(x_{2})$} q[2];
            [name=e-bot]
            box {$R_x(x_{3})$} q[3];

            hspace {5mm} -;
            [name=v-top]
            box {$R_y(\theta_{0})$} q[0];
            box {$R_y(\theta_{1})$} q[1];
            box {$R_y(\theta_{2})$} q[2];
            [name=v-bot]
            box {$R_y(\theta_{3})$} q[3];

            box {$R_z(\theta_{4})$} q[0];
            box {$R_z(\theta_{5})$} q[1];
            box {$R_z(\theta_{6})$} q[2];
            box {$R_z(\theta_{7})$} q[3];

            zz (q[3, 2]);
            zz (q[2, 1]);
            [name=v-right]
            zz (q[1, 0]);
            
            align -;
            hspace {5mm} -;
            
            [name=d-top]
            output {$Z$} q[0];
            output {$Z$} q[1];
            output {$Z$} q[2];
            [name=d-bot]
            output {$Z$} q[3];
            
        \end{yquant}
        \node[draw, dashed, fit=(e-top) (e-bot), label=Encoding] {};
        \node[draw, dashed, fit=(v-top) (v-bot) (v-top) (v-right), label=Variational] {};
        \node[draw, dashed, fit=(d-top) (d-bot), label=Decoding] {};
    \end{tikzpicture}
    }
    \caption{A simple variational quantum circuit depicting an encoding layer, variational layer and a decoding layer.}
    \label{fig:VQC}
\end{figure}
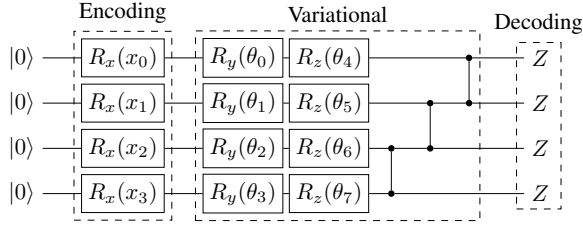

A \gls{VQC} is formally described as a sequence of parameterized gates $U(\bm{x, \theta})$ acting on an initial quantum state $|\psi_0\rangle$ which is often but not necessarily a $|0\rangle$ state, input vector $\bm{x}$ and variational parameter vector $\bm{\theta}$. A typical \gls{VQC} consists of three layers, namely the encoding layer, the variational layer, and the decoding layer. The angle encoding method is the most prevalent encoding strategy. Here, the classical data is encoded using the Pauli rotational gates. The angle encoding also allows for some advanced variations of encoding, such as the data re-uploading scheme \cite{Salinas2020} and the cyclic data re-uploading scheme~\cite{periyasamy2024bcqq}. The variation layer is typically made of different Pauli rotational gates and multi-qubit entanglement gates. The decoding layer is the decoding technique used, such as expectation value with respect to an observable. \cref{fig:VQC} represents the structure of a \gls{VQC}.

The unitary operation $U(\bm{x, \theta})$ transforms the initial state $|\psi_0\rangle$ to a new state $|\psi(\bm{\theta})\rangle$ by

\begin{equation}
|\psi(\bm{\theta})\rangle = U(\bm{x, \theta})|\psi_0\rangle.
\end{equation}

\subsubsection{Objective Function}

In QML, the goal is to find the set of parameters $\bm{\theta}$ that minimize a given objective function. The output function  $f(\bm{x, \theta}))$ is defined with respect to some observable $A$ as shown in \cref{eq:expectation_value}. The loss function $L(f(\bm{x, \theta}), \bm{y})$ is defined based on the type of the problem and the target values $y$.

\begin{equation}
\label{eq:expectation_value}
f(\bm{x, \theta})) = \langle \psi(\bm{\theta}) | A | \psi(\bm{\theta}) \rangle
\end{equation}

\subsection*{Optimization}

To find the optimal parameters $\bm{\theta}^*$ that minimize the objective function $L(f(\bm{x, \theta}), \bm{y})$, optimization algorithms such as gradient descent are commonly employed. The parameters are iteratively updated according to the gradient of the objective function:

\begin{equation}
\label{eq:gradient_descent}
\bm{\theta}^{(t+1)} = \bm{\theta}^{(t)} - \eta \nabla L(\bm{\theta}^{(t)})
\end{equation}

where $\eta$ denotes the learning rate and $\nabla L(\bm{\theta})$ is the gradient of the objective function with respect to the parameters.

\subsection{Gradient Computation}
\label{sec:gradient_estimation}
Performing the optimization step in \cref{eq:gradient_descent} requires access to the gradient \(\nabla L(\bm{\theta})\) of the objective function.
The chain rule allows us to determine the gradient \(\nabla L(\bm{\theta})\), once the partial derivatives of the output function with respect to the variational parameters $\partial_{\theta_i} f(\bm{\theta})$ is known. The following two subsections discuss two state-of-the-art approaches to compute these partial derivatives.

\subsubsection{Parameter-shift rule}

The parameter-shift rule was developed by \cite{Schuld19, Crooks19} to obtain exact expressions for the partial derivatives $\partial_{\theta_i} f(\bm{\theta})$ from expectation values that can be estimated on the quantum hardware. The parameter-shift rule states that the partial derivative for parameter \(i\) is exactly given by
\begin{equation}
\label{eq:param_shift}
    \partial_{\theta_i} f(\bm{\theta}) = r \left[ f(\bm{\theta} + \bm{s}) -  f(\bm{\theta} - \bm{s})\right],
\end{equation}
where $r$ depends on the two distinct eigenvalues of the hermitian generator of the circuit and  $\bm{s}$ is the shift value. With the parameter-shift rule, the $\partial_{\theta_i} f(\bm{\theta})$ can be estimated on the quantum hardware by estimating two expectation values of the same quantum circuit with shifted parameters. Therefore, for the whole gradient, one needs to estimate twice as many expectation values as parameters in the circuit.

\subsubsection{Stochastic Perturbation Simultaneous Approximation}
\label{sec:spsa}
In contrast to the gradient based optimization routines, \gls{SPSA} \cite{Spall92} does not require access to estimates of the partial derivatives. Moreover, it was specifically designed for situations in which the objective function evaluation is corrupted by noise. \Gls{SPSA} uses a stochastic approximation instead of the actual gradient of the cost function in the parameter update step given by \cref{eq:gradient_descent}. On average, over many perturbation samples, \gls{SPSA} will eventually follow the direction of the steepest descent.
The approximation $\hat{\bm{g}}^k(\hat{\bm{\theta}}^k) \approx \nabla L(\hat{\bm{\theta}}^k)$ is computed as
\begin{equation*}
\label{eq:spsa}
    \hat{\bm{g}}^k(\hat{\bm{\theta}}^k) = \frac{L(\hat{\bm{\theta}}^k + c^k\bm{\Delta}^k) - L(\hat{\bm{\theta}}^k - c^k\bm{\Delta}^k)}{2c^k} \begin{bmatrix}
	(\Delta^k_1)^{-1} \\
	(\Delta^k_2)^{-1} \\
	\vdots \\
	(\Delta^k_p)^{-1}\\
	\end{bmatrix},
\end{equation*}
Where $c$ corresponds to the size and $\bm{\Delta}$ to the random direction of a perturbation similar to the finite difference method of numerical derivation, the upper index \(k\) denotes the perturbation sample. The entries $\Delta^k_i$ of the perturbation direction are sampled independently from a uniform distribution over the set \(\{-1, 1\}\).

The key difference to the parameter-shift rule is that all parameters are perturbed at the same time. Therefore, the whole update step always requires only exactly two expectation value estimations times the number of perturbation samples used. This can drastically speed up the training process as long as the number of update steps required to reach the optimum does not increase significantly. Theoretical and empirical considerations confirm that this is not the case \cite{Kungurtsev2022}.

\subsection{Training Algorithm for Quantum Regression}
\label{sec:Optimizers}
\begin{algorithm}[tbp!]
\SetAlgoLined
\KwIn{Training data: $\mathcal{D} = (X, y)$, Number of epochs: $N_{\text{epochs}}$, Batch size: $B$, Learning rate: $\eta$,  \gls{VQC}: $U(x, \vec{\theta})$}
\KwOut{Optimized parameters: $\vec{\theta}$\;}
Initialize parameters $\vec{\theta}$ randomly\;
Normalize dataset \(\mathcal{D}\) between \([-\pi, \pi]\)

\For{$epoch = 0$ \KwTo $N_{\text{epochs}} - 1$}{
    \For{$batch = 0$ \KwTo $ \frac{length(D)}{B}-1$}{
        Sample mini-batch  $M = (X_{\text{batch}}, y_{\text{batch}}) \in \mathcal{D}$ of size $B$\;
        \SetKwProg{FP}{Forward pass}{:}{}
        \FP{}{
        Compute predictions $\hat{y}_{\text{batch}} = \langle \psi(\vec{\theta}) | A | \psi(\vec{\theta}) \rangle$\;
        Compute error $\vec{E}$ using $(y_{\text{batch}}, \hat{y}_{\text{batch}})$\;
        }
        \SetKwProg{BP}{Backward pass}{:}{}
        \BP{}{
        Compute gradients $\nabla_{\vec{\theta}}$\; 
        Update parameters using chosen optimizer:
        $\vec{\theta} = optimizer(\vec{\theta}, \eta, \nabla_{\vec{\theta}}\vec{E})$\;
        }
    }
    \If{stopping criteria met}{
    break \;
  }
}
\caption{Quantum Machine Learning}
\label{alg:training_vqc}
\end{algorithm}

\Gls{QML}, like its classical counterpart, has multiple branches such as regression, classification, and reinforcement learning. Quantum regressors are models where a \gls{VQC} predicts one or more continuous values for a given input \cite{Wiedmann2023}. Quantum classifiers are function approximators that predict the probability of a given input belonging to a specific class \cite{Periyasamy2022, Senokosov2024, John2023}. Quantum reinforcement learning (QRL) is a type of machine learning technique that enables an agent to learn through trial and error within an environment. Quantum versions of classical reinforcement learning such as policy gradient method \cite{Meyer_2022a, Meyer_2023a, Meyer_2023b}, DQN \cite{franz2022uncovering, Skolik2021, Chen_2022}, and offline RL \cite{periyasamy2024bcqq} have been studied in the past.

Training a \gls{VQC} $U(\bm{x, \theta})$ for regression or any learning task, in general, involves feeding it a normalized dataset and comparing its predictions with the actual target values. An error function such as mean squared error, mean absolute error, etc., quantifies this difference. An optimizer function like SGD\cite{sutskever2013importance}, ADAM\cite{ADAM}, AMSGrad\cite{Reddi19}, RMSProp\cite{tieleman2012lecture}, etc., tunes the trainable parameters. Hyperparameters, separate from the \gls{VQC} architecture, control this optimization process. Tuning these hyperparameters significantly impacts the model's performance. \cref{alg:training_vqc} shows a simple quantum regression training procedure.  The training procedure is a quantum-classical hybrid process where the quantum part of the algorithm is also often simulated on a classical device using a quantum simulator due to the limited availability of NISQ devices. There are two different categories of quantum simulators, namely, an ideal simulator that calculates the exact expectation values and a shot-based simulator that approximates the expectation value using finite samples. Further, one can perform a noisy quantum simulation on a classical device that replicates the properties of the NISQ hardware, such as limited qubit connectivity, decoherence, gate infidelity, and crosstalk to a certain degree using the shot-based simulator and a noise model. Quantum error mitigation (QEM) \cite{Cai22} approaches such as zero-noise extrapolation (ZNE) \cite{temme2017error} can also be integrated into a noisy simulator, which aids in the development of quantum algorithms tailored towards the current capabilities of the hardware. Methods like ZNE attempt to extrapolate the behavior of the circuit in the ideal, noiseless limit. This extrapolation often relies on polynomial fitting or other statistical methods. As the hardware noise and the increase in the number of circuit evaluations introduced by ZNE influence the stability of the training procedure, testing the Guided-SPSA under these conditions is of prime interest to us.

\begin{algorithm*}[htbp!]
\SetAlgoLined
\KwIn{Training data: $\mathcal{D} = (X, y)$, Number of epochs: $N_{\text{epochs}}$, Batch size: $B$, Learning rate: $\eta$,  \gls{VQC}: $U(x, \vec{\theta})$, Parameter-shift sample ratio: $\tau$, SPSA damping constant: $\epsilon$\;}
\KwOut{Optimized parameters: $\vec{\theta}$\;}
Initialize parameters $\vec{\theta}$ randomly\;

Set number of batches $m = \frac{length(\mathcal{D})}{B}$\;
Normalize dataset \(\mathcal{D}\) between \([-\pi, \pi]\)

Set maximum number of SPSA perturbation samples $k_{\text{max}} = length(\vec{\theta}) \times min(1, 1.5-\tau) $ \;
Set minimum number of SPSA perturbation samples $k_{\text{min}} =  max(1, length(\vec{\theta}) \times 0.1)$ \;
Calculate SPSA sample increment factor $\gamma =  \frac{k_{\text{max}} - k_{\text{min}}}{N_{\text{epochs}}}$ 

\For{$epoch = 0$ \KwTo $N_{\text{epochs}} - 1$}{
    Set current SPSA perturbation sample $k_{\text{epoch}} = \floor*{k_{\text{min}} + epoch*\gamma}$\;
    Shuffle training data $\mathcal{D}$\;
    
    \For{$batch = 0$ \KwTo $m-1$}{
        Sample mini-batch  $M = (X_{\text{batch}}, y_{\text{batch}}) \in \mathcal{D}$ of size $B$\;
        \SetKwProg{FP}{Forward pass}{:}{}
        \FP{}{
        Compute predictions $\hat{y}_{\text{batch}} = \langle \psi(\vec{\theta}) | A | \psi(\vec{\theta}) \rangle$\;
        Compute error $\vec{E}$ using $(y_{\text{batch}}, \hat{y}_{\text{batch}})$\;
        }
        \SetKwProg{BP}{Backward pass}{:}{}
        \BP{}{
        Split $M-> M_\text{ps}, M_\text{spsa}$ where $length(M_\text{ps}) = \tau \times B, length(M_\text{spsa}) = (1-\tau) \times B$\;
        Calculate $\nabla_\text{ps}$ for each data in $M_\text{ps}$ using parameter shift rule\;
        Calculate average norm per observable of parameter-shift gradients $\sigma = \frac{1}{\tau \times B}\sum\limits_{i=1}^{\tau \times B}{||\nabla_\text{ps}^i||_2}$

        \For{$each\ m$ in $M_\text{spsa}$}{
        Calculate $\nabla_\text{spsa}^{m}$ for $m$ using SPSA based gradient estimation\;
        Suppress the gradient length of $\nabla_\text{spsa}^{m}$ as follows:
            $\nabla_\text{g-spsa}^{m} =  \frac{\sigma}{||\nabla_\text{spsa}^{m}||_2}\times \epsilon \times \nabla_\text{spsa}^{m}$
        }

        $\nabla_{\vec{\theta}} = \begin{bmatrix}
            \nabla_\text{ps} \\
            \nabla_\text{g-spsa}
        \end{bmatrix}$
        Update parameters using chosen optimizer:
        $\vec{\theta} = optimizer(\vec{\theta}, \eta, \nabla_{\vec{\theta}}\vec{E})$\;
        
        }
    }
}
\caption{Guided-SPSA}
\label{alg:guided-spsa}
\end{algorithm*}

\section{Guided-SPSA}
\label{sec:guided-spsa}

As mentioned in \cref{sec:gradient_estimation}, the parameter-shift rule estimates the exact gradients with respect to parameters. However, the parameter shift rule has a computational complexity of $\mathcal{O}(2NM)$ where $M$ is the number of parameters and $N$ is the number of training data points. This dependence on the number of parameters $M$ is undesirable as the computation complexity increases for larger models; additionally, it is not well suited for NISQ devices because of noise and imperfections. The SPSA-based gradient approximation computes approximate gradients with a computational complexity of $\mathcal{O}(2kM)$ where $k$ is the number of perturbation samples; this number is a design choice. The complexity of SPSA-based gradient estimation does not necessarily grow with the number of circuit parameters, but the accuracy of the gradient is directly proportional to $k$. In this work, we leverage the advantages of both the parameter-shift rule and SPSA by splitting the input samples between both methods for gradient estimation and meaningfully combining the gradients during the training process. 

The Guided-SPSA algorithm is presented in \cref{alg:guided-spsa}. The Guided-SPSA introduces two new hyperparameters as design choices, namely, parameter-shift sample ratio $\tau$ and \gls{SPSA} damping constant $\epsilon$. Parameter-shift sample ratio $\tau$ defines the ratio of the input samples for which the parameter shift rule is to be used for gradient estimation. The remaining samples will use the SPSA method. Parameter-shift sample ratio $\tau$ should be defined between $[0,1]$ where $\tau = 1$ is the standard parameter-shift rule procedure and $\tau=0$ implies standard SPSA procedure. \gls{SPSA} damping constant $\epsilon$ is used to suppress the magnitude of the gradients from \gls{SPSA} method. $\epsilon$ ranges between $(0, 1]$ where the value of 1 means that the gradients from \gls{SPSA} is not manually suppressed. $\epsilon$ close to 0 leads to heavy suppression of the magnitude. \gls{SPSA} damping constant $\epsilon$ helps to attain a stable convergence when the training data is noisy and contains distribution samples. The algorithm starts by receiving the following inputs: training data $\mathcal{D}$, number of epochs $N_{\text{epochs}}$, batch size $B$, learning rate $\eta$,  the \gls{VQC} ansatz $U(x, \vec{\theta})$, $\tau$, and $\epsilon$. First, we calculate the maximum and minimum number of perturbation sample $k$ to be used during the training procedure based on the number of parameters in the \gls{VQC} and $\tau$ as shown in step 5, step 6 in \cref{alg:guided-spsa}. The perturbation samples for \gls{SPSA} start from $k_{min}$ and increase linearly to $k_{max}$ during the course of the training using $\gamma$. Here, we leverage the advantage of the SPSA-based gradient estimation by using a smaller perturbation sample size $k$ at the start of the training and linearly increase it as the training progresses. These small perturbation sample sizes drastically reduce the computational complexity of the overall gradient estimation during the early stages of the training at the cost of introducing some inaccuracies in the gradients. For every epoch and each batch in those epochs, the forward pass resembles the standard training procedure as shown in \cref{alg:training_vqc}. 

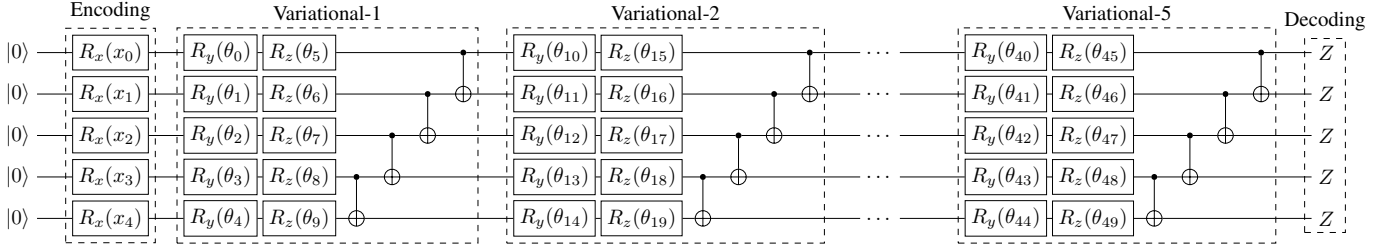
\begin{figure*}[htbp!]
    \centering
    \begin{tikzpicture}
        \begin{yquant}
            qubit {$\ket0$} q[5];
            hspace {5mm} -;
            [name=e-top]
            box {$R_x(x_{0})$} q[0];
            box {$R_x(x_{1})$} q[1];
            box {$R_x(x_{2})$} q[2];
            box {$R_x(x_{3})$} q[3];
            [name=e-bot]
            box {$R_x(x_{4})$} q[4];

            hspace {5mm} -;
            [name=v-top]
            box {$R_y(\theta_{0})$} q[0];
            box {$R_y(\theta_{1})$} q[1];
            box {$R_y(\theta_{2})$} q[2];
            box {$R_y(\theta_{3})$} q[3];
            [name=v-bot]
            box {$R_y(\theta_{4})$} q[4];

            box {$R_z(\theta_{5})$} q[0];
            box {$R_z(\theta_{6})$} q[1];
            box {$R_z(\theta_{7})$} q[2];
            box {$R_z(\theta_{8})$} q[3];
            box {$R_z(\theta_{9})$} q[4];

            cnot q[4] | q[3];
            cnot q[3] | q[2];
            cnot q[2] | q[1];
            [name=v-right]
            cnot q[1] | q[0];
            
            align -;
            hspace {5mm} -;
            
            [name=v1-top]
            box {$R_y(\theta_{10})$} q[0];
            box {$R_y(\theta_{11})$} q[1];
            box {$R_y(\theta_{12})$} q[2];
            box {$R_y(\theta_{13})$} q[3];
            [name=v1-bot]
            box {$R_y(\theta_{14})$} q[4];

            box {$R_z(\theta_{15})$} q[0];
            box {$R_z(\theta_{16})$} q[1];
            box {$R_z(\theta_{17})$} q[2];
            box {$R_z(\theta_{18})$} q[3];
            box {$R_z(\theta_{19})$} q[4];

            cnot q[4] | q[3];
            cnot q[3] | q[2];
            cnot q[2] | q[1];
            [name=v1-right]
            cnot q[1] | q[0];

            align -;
            
            hspace {5mm} -;
            
            text {$\dots$} q;

            align -;
            hspace {5mm} -;
            


            
            



            align -;

            hspace {5mm} -;
            
            [name=v4-top]
            box {$R_y(\theta_{40})$} q[0];
            box {$R_y(\theta_{41})$} q[1];
            box {$R_y(\theta_{42})$} q[2];
            box {$R_y(\theta_{43})$} q[3];
            [name=v4-bot]
            box {$R_y(\theta_{44})$} q[4];

            box {$R_z(\theta_{45})$} q[0];
            box {$R_z(\theta_{46})$} q[1];
            box {$R_z(\theta_{47})$} q[2];
            box {$R_z(\theta_{48})$} q[3];
            box {$R_z(\theta_{49})$} q[4];

            cnot q[4] | q[3];
            cnot q[3] | q[2];
            cnot q[2] | q[1];
            [name=v4-right]
            cnot q[1] | q[0];
            
            align -;
            hspace {5mm} -;
            
            [name=d-top]
            output {$Z$} q[0];
            output {$Z$} q[1];
            output {$Z$} q[2];
            output {$Z$} q[3];
            [name=d-bot]
            output {$Z$} q[4];
            
        \end{yquant}
        \node[draw, dashed, fit=(e-top) (e-bot), label=Encoding] {};
        \node[draw, dashed, fit=(v-top) (v-bot) (v-top) (v-right), label=Variational-1] {};
        \node[draw, dashed, fit=(v1-top) (v1-bot) (v1-top) (v1-right), label=Variational-2] {};
        \node[draw, dashed, fit=(v4-top) (v4-bot) (v4-top) (v4-right), label=Variational-5] {};
        
        \node[draw, dashed, fit=(d-top) (d-bot), label=Decoding] {};
    \end{tikzpicture}
    
    \caption{VQC used for 5 qubit circut experiments such as Friedman regression}
    \label{fig:VQC_5qubit}
\end{figure*}

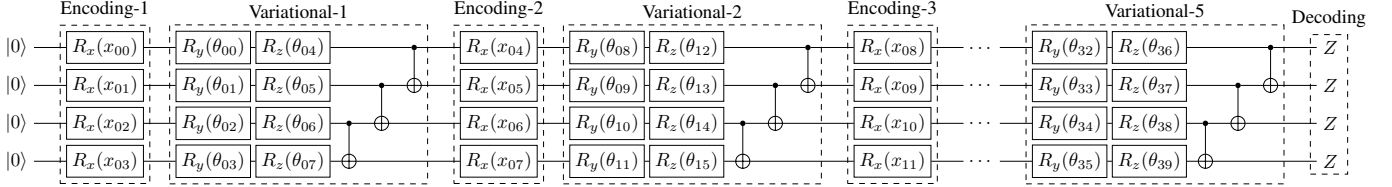
\begin{figure*}[htbp!]
    \centering
    \begin{tikzpicture}
        \begin{yquant}
            qubit {$\ket0$} q[4];
            hspace {5mm} -;
            [name=e-top]
            box {$R_x(x_{00})$} q[0];
            box {$R_x(x_{01})$} q[1];
            box {$R_x(x_{02})$} q[2];
            [name=e-bot]
            box {$R_x(x_{03})$} q[3];
            
            hspace {5mm} -;
            [name=v-top]
            box {$R_y(\theta_{00})$} q[0];
            box {$R_y(\theta_{01})$} q[1];
            box {$R_y(\theta_{02})$} q[2];
            [name=v-bot]
            box {$R_y(\theta_{03})$} q[3];
            
            box {$R_z(\theta_{04})$} q[0];
            box {$R_z(\theta_{05})$} q[1];
            box {$R_z(\theta_{06})$} q[2];
            box {$R_z(\theta_{07})$} q[3];
            
            cnot q[3] | q[2];
            cnot q[2] | q[1];
            [name=v-right]
            cnot q[1] | q[0];
            
            align -;
            hspace {5mm} -;
            
            [name=e1-top]
            box {$R_x(x_{04})$} q[0];
            box {$R_x(x_{05})$} q[1];
            box {$R_x(x_{06})$} q[2];
            [name=e1-bot]
            box {$R_x(x_{07})$} q[3];
            
            hspace {5mm} -;

            [name=v1-top]
            box {$R_y(\theta_{08})$} q[0];
            box {$R_y(\theta_{09})$} q[1];
            box {$R_y(\theta_{10})$} q[2];
            [name=v1-bot]
            box {$R_y(\theta_{11})$} q[3];

            box {$R_z(\theta_{12})$} q[0];
            box {$R_z(\theta_{13})$} q[1];
            box {$R_z(\theta_{14})$} q[2];
            box {$R_z(\theta_{15})$} q[3];
            
            cnot q[3] | q[2];
            cnot q[2] | q[1];
            [name=v1-right]
            cnot q[1] | q[0];
            
            align -;
            hspace {5mm} -;

            [name=e2-top]
            box {$R_x(x_{08})$} q[0];
            box {$R_x(x_{09})$} q[1];
            box {$R_x(x_{10})$} q[2];
            [name=e2-bot]
            box {$R_x(x_{11})$} q[3];
            
            hspace {5mm} -;
            
            text {$\dots$} q;
            
            hspace {5mm} -;
            
            
           
            

            


            
            

            
            [name=v4-top]
            box {$R_y(\theta_{32})$} q[0];
            box {$R_y(\theta_{33})$} q[1];
            box {$R_y(\theta_{34})$} q[2];
            [name=v4-bot]
            box {$R_y(\theta_{35})$} q[3];

            box {$R_z(\theta_{36})$} q[0];
            box {$R_z(\theta_{37})$} q[1];
            box {$R_z(\theta_{38})$} q[2];
            box {$R_z(\theta_{39})$} q[3];
           
            cnot q[3] | q[2];
            cnot q[2] | q[1];
            [name=v4-right]
            cnot q[1] | q[0];
            
            align -;
            hspace {5mm} -;
            
            [name=d-top]
            output {$Z$} q[0];
            output {$Z$} q[1];
            output {$Z$} q[2];
            [name=d-bot]
            output {$Z$} q[3];
           
        \end{yquant}
        \node[draw, dashed, fit=(e-top) (e-bot), label=Encoding-1] {};
        \node[draw, dashed, fit=(e1-top) (e1-bot), label=Encoding-2] {};
        \node[draw, dashed, fit=(e2-top) (e2-bot), label=Encoding-3] {};
        \node[draw, dashed, fit=(v-top) (v-bot) (v-top) (v-right), label=Variational-1] {};
        \node[draw, dashed, fit=(v1-top) (v1-bot) (v1-top) (v1-right), label=Variational-2] {};
        \node[draw, dashed, fit=(v4-top) (v4-bot) (v4-top) (v4-right), label=Variational-5] {};
        
        \node[draw, dashed, fit=(d-top) (d-bot), label=Decoding] {};
    \end{tikzpicture}
    \caption{VQC used for 4 qubit circut experiments such as toy problem and boston-housing regression}
    \label{fig:VQC_4qubit}
\end{figure*}

During the backward pass, the batch samples $M$ is split into to parts, namely, $M_{\text{ps}}$ and $M_{\text{spsa}}$ based on $\tau$ first. Then, the gradient of the parameter vector $\theta$ for the samples $M_{\text{ps}}$ are calculated using the parameter-shift rule. Next, we do the same for the samples $M_{\text{spsa}}$ but using \gls{SPSA} rule and each gradient vector is suppressed using $\epsilon$ and $\sigma$ as shown in step 18 to step 22. Here, we leverage the advantage of the parameter-shift rule, that is, the exactness of the derivatives, and use it to guide the derivatives calculated by SPSA. By adjusting the magnitude of the SPSA with the magnitude of the gradients from the parameter-shift rule, we reduce the chances of the solution moving too far from the exact gradients. This reduces the instability, especially during the early stage of the training when the SPSA perturbation sample is too small and the last stages of the training when the solution is potentially in the neighborhood of the minima. Finally, the combined gradients are multiplied with the error vector $\vec{E}$ to form the gradient with respect to the loss based on the chain rule.

\section{Experiments and Results}
\label{sec:Results}

In this section, we present the results of the numerical experiment conducted to empirically validate the performance of Guided-SPSA. First, we look at a toy problem where we minimize a simple function using the Guided SPSA. Then, we examine the performance using various regression tasks under different experimental conditions. Finally, we test the algorithm on a classification and reinforcement learning problem.

\subsection{Toy Problem}
\label{sec:toy_problem}

\begin{figure}[htbp]
  \vspace*{-1em}\input{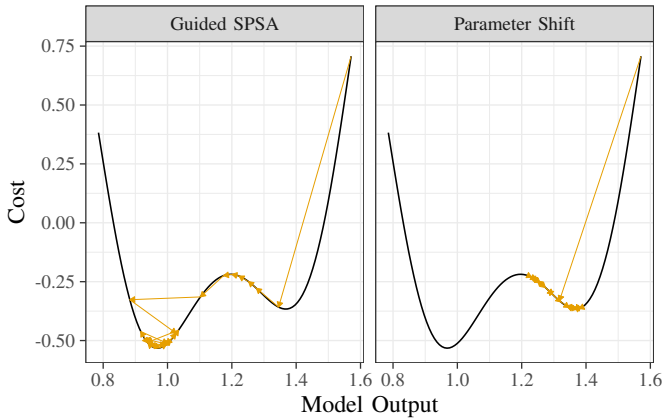}\vspace*{-2em}
     \caption{Path for finding minima of the cost function given in \cref{eq:toy_problem} (black) using parameter-shift rule and Guided-\gls{SPSA}} \label{fig:toy_param_shift}
\end{figure}

To better visualize and understand the convergence of parameter-shift-based gradients and Guided-SPSA, we utilized respective gradient estimation techniques to minimize the function specified in \cref{eq:toy_problem}. The task of the VQC is to find the value of $x$ for which the \cref{eq:toy_problem} is minimized. The toy problem \cref{eq:toy_problem} was chosen as it has
a local minimum, cf. \cref{fig:toy_param_shift}, which the algorithm should be able to avoid.

\begin{equation}
    \label{eq:toy_problem}
    L(x) = \sin\left(\frac{x}{2}\right) + \sin(2.25 \cdot \sin(4 \cdot x)) 
\end{equation}

\cref{fig:toy_param_shift} shows the convergence of the \cref{eq:toy_problem} to a minima using parameter-shift rule and Guided-\gls{SPSA}. We used the VQC shown in \cref{fig:VQC_4qubit} without any encoding layer as the function approximator for both experiments. The encoding layer was unnecessary as there is no classical input to the VQC in this problem case. All the hyper-parameters, such as learning rate, number of epochs, etc., were the same for both runs. \cref{fig:toy_param_shift} shows that the Guided-\gls{SPSA} escaped the local minima compared to the parameter-shift rule. This is due to the randomness and inaccuracies brought in by the SPSA components at the early stage of the training.

\subsection{Regression}

\begin{table*}[btp]
\caption{Regression results trained using different simulators}
\label{tab:regression}
\centering

\begin{tabular}{ccccc|cccc}
\toprule
Method                                                      & No. Circuits                                                          & Convergence   & Validation    & Test          & No. Circuits                                                          & Convergence   & Validation    & Test          \\
\midrule
                                                            & \multicolumn{4}{c}{Friedman}                                                                                          & \multicolumn{4}{c}{Boston Housing}                                                                                    \\
\cmidrule{2-9}
\addlinespace
\multicolumn{9}{c}{Ideal simulation}                                                                                                                                                                                                                                                                        \\
\addlinespace

SPSA-10                                                     & 1M                                                                    & $\sim$77      & 0.08          & 0.10          & 0.75M                                                                 & $\sim$63      & 0.19          & 0.21          \\
SPSA-20                                                     & 2M                                                                    & $\sim$77      & 0.07          & 0.09          & 1.45M                                                                 & $\sim$54      & 0.18          & 0.19          \\
SPSA-30                                                     & 3M                                                                    & $\sim$83      & 0.07          & 0.08          & 2.17M                                                                 & $\sim$67      & 0.17          & 0.19          \\
SPSA-max                                                    & 5M                                                                    & $\sim$85      & 0.06          & 0.08          & 2.90M                                                                 & $\sim$74      & 0.17          & 0.19          \\
\begin{tabular}[c]{@{}c@{}}Parameter-\\ Shift\end{tabular}  & 5M                                                                    & $\sim$75      & 0.06          & 0.07          & 2.90M                                                                 & $\sim$63      & \textbf{0.16} & \textbf{0.18} \\
Guided-SPSA                                                 & \begin{tabular}[c]{@{}c@{}}3.84M\\(\textbf{$\sim$24\%})\end{tabular}  & $\sim$83      & \textbf{0.05} & \textbf{0.06} & \begin{tabular}[c]{@{}c@{}}2.38M\\(\textbf{$\sim$18\%})\end{tabular}  & $\sim$65      & \textbf{0.16} & \textbf{0.18} \\

\addlinespace
\multicolumn{9}{c}{Shot based}                                                                                                                                                                                                                                                                              \\
\addlinespace

SPSA-10                                                     & 1M                                                                    & $\sim$68      & 0.11          & 0.15          & 0.75M                                                                 & $\sim$63      & 0.19          & 0.21          \\
SPSA-20                                                     & 2M                                                                    & $\sim$77      & 0.08          & 0.11          & 1.45M                                                                 & $\sim$65      & 0.17          & 0.20          \\
SPSA-30                                                     & 3M                                                                    & $\sim$79      & 0.07          & 0.09          & 2.17M                                                                 & $\sim$63      & 0.17          & 0.20          \\
SPSA-max                                                    & 5M                                                                    & $\sim$76      & 0.07          & \textbf{0.08} & 2.90M                                                                 & $\sim$71      & 0.17          & 0.19          \\
\begin{tabular}[c]{@{}c@{}}Parameter-\\ Shift\end{tabular}  & 5M                                                                    & $\sim$75      & \textbf{0.06} & \textbf{0.08} & 2.90M                                                                 & $\sim$68      &\textbf{0.16}  & \textbf{0.18} \\
Guided-SPSA                                                 & \begin{tabular}[c]{@{}c@{}}3.84M\\(\textbf{$\sim$24\%})\end{tabular}  & $\sim$78      & \textbf{0.06} & \textbf{0.08} & \begin{tabular}[c]{@{}c@{}}2.38M\\(\textbf{$\sim$18\%})\end{tabular}  & $\sim$67      &\textbf{0.16}  & \textbf{0.18} \\

\addlinespace
\multicolumn{9}{c}{Noisy Simulation}                                                                                                                                                                                                                                                                        \\
\addlinespace

SPSA-10                                                     & 1M                                                                    & $\sim$69      & 0.11          & 0.14          & 0.75M                                                                 & $\sim$71      & 0.18          & 0.21          \\
SPSA-20                                                     & 2M                                                                    & $\sim$77      & 0.08          & 0.10          & 1.45M                                                                 & $\sim$78      & 0.17          & 0.20          \\
SPSA-30                                                     & 3M                                                                    & $\sim$74      & 0.07          & 0.09          & 2.17M                                                                 & $\sim$63      & 0.17          & \textbf{0.19} \\
SPSA-max                                                    & 5M                                                                    & $\sim$74      & 0.07          & 0.09          & 2.90M                                                                 & $\sim$79      & 0.17          & \textbf{0.19} \\
\begin{tabular}[c]{@{}c@{}}Parameter-\\ Shift\end{tabular}  & 5M                                                                    & $\sim$73      & \textbf{0.06} & \textbf{0.08} & 2.90M                                                                 & $\sim$62      & \textbf{0.16} & \textbf{0.19} \\
Guided-SPSA                                                 & \begin{tabular}[c]{@{}c@{}}3.84M\\(\textbf{$\sim$24\%})\end{tabular}  & $\sim$80      & \textbf{0.06} & \textbf{0.08} & \begin{tabular}[c]{@{}c@{}}2.38M\\(\textbf{$\sim$18\%})\end{tabular}  & $\sim$60      & \textbf{0.16} & \textbf{0.19} \\

\addlinespace
\multicolumn{9}{c}{Error Mitigation}                                                                                                                                                                                                                                                                        \\
\addlinespace

\begin{tabular}[c]{@{}c@{}}Parameter-\\ Shift\end{tabular} & 5M                                                                     & $\sim$77      & \textbf{0.06} & \textbf{0.08} & 2.90M                                                                 & $\sim$69      & \textbf{0.16} & \textbf{0.18} \\
Guided-SPSA                                                & \begin{tabular}[c]{@{}c@{}}3.84M\\(\textbf{$\sim$24\%})\end{tabular}   & $\sim$79      & \textbf{0.06} & \textbf{0.08} & \begin{tabular}[c]{@{}c@{}}2.38M\\(\textbf{$\sim$18\%})\end{tabular}  & $\sim$63      & \textbf{0.16} & \textbf{0.18} \\

\bottomrule
\end{tabular}%
\end{table*}

\subsubsection{Datasets}
\label{sec:datasets}
In order to study the performance of different gradient estimation techniques in training a VQC beyond the toy problem, we conducted a series of regression training runs. Here, we trained the VQC with the same initial parameters and ansatz using SPSA of various perturbation sizes, parameter-shift rule, and Guided-SPSA as defined in \cref{alg:guided-spsa}. The following datasets were employed in the experiments: 
\begin{enumerate*}
    \item Friedman dataset: A five-dimensional dataset, first introduced by Friedman \cite{friedman1991multivariate, breiman1996bagging}. This data set is constructed from rational and trigonometric functions of the input features. We sampled a total of 500 points from this dataset.
    \item  Boston-housing dataset: A fourteen-dimensional dataset consisting of various features around a housing community in Boston, which can be used to predict the nitrous oxide level based on the remaining thirteen features \cite{harrub78}. This dataset consists of 506 samples in total.
\end{enumerate*}
Further, all the features in the dataset were normalized between $[-\pi, \pi]$ to encode into the VQC. For all training runs, we used an approximate train/validation/test sample ratio of 0.68/0.22/0.1.

\subsubsection{Problem Setup and Hyperparameters}
\label{sec:hyperparameters}
The function approximator used for these regression tasks consists of four or five-qubit circuits. The VQC shown in \cref{fig:VQC_5qubit} was used for all the Friedman dataset experiments. This VQC consists of one encoding layer and five variational layers. The encoding layer is made of single qubit Pauli-X rotational gates, and the variational layer consists of Pauli-Y and Pauli-Z rotational gates. Furthermore, we have used CNot entanglement with nearest-neighbour connectivity in all variational layers. The output is decoded using the observable \(A = Z^{\otimes n}\) as shown in \cref{eq:expectation_value}. Likewise, the VQC shown in \cref{fig:VQC_4qubit} was used for all the Boston housing dataset experiments. This VQC is made of a four-qubit circuit and is similar to the VQC used for Friedman dataset experiments in terms of the gates used. However, to encode a thirteen-dimensional feature vector into a four-qubit circuit, we employ the incremental data-uploading technique introduced by Periyasamy~\etal \cite{Periyasamy2022}. We use the Adam optimizer with a learning rate of 0.01, MSE as the loss function, and MAE as the accuracy metric for all the experiments.

\subsubsection{Ideal Simulation}
\label{sec:ideal_simulation}
For the first set of experiments, we train the VQCs for the regression task explained in \cref{sec:datasets} and \cref{sec:hyperparameters} using an exact quantum simulator. Here, the simulation mimics ideal quantum hardware with all-to-all qubit connectivity and does not include the influence of shot noise. To form equal test conditions and attain statistical veracity, we sampled five sets of initial parameters from a uniform distribution between $[0, \pi]$. We trained all the methods five times with each parameter set. A total of twenty-five experiments per gradient estimation method were performed, and the average results over these twenty-five runs are presented.

\begin{figure}[tbp!]
  \vspace*{-1em}\input{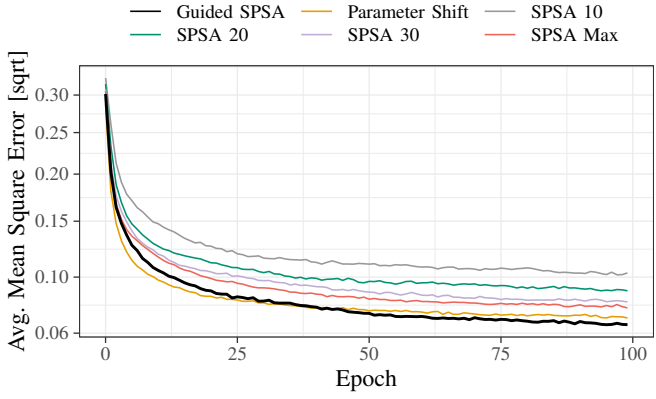}\vspace*{-2em}
     \caption{Convergence of different gradient estimation techniques using an ideal simulator on the Friedman dataset.}
    \label{fig:train_friedmaan_ideal}
\end{figure}

\begin{figure}[tbp!]
  \vspace*{-1em}\input{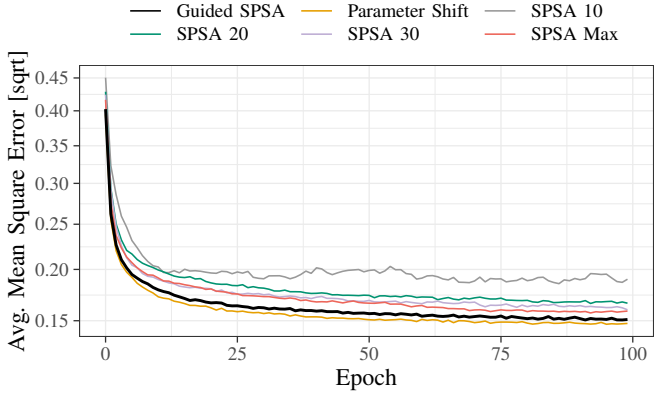}\vspace*{-2em}
     \caption{Convergence of different gradient estimation techniques using an ideal simulator on the Boston housing dataset.}
    \label{fig:train_boston_ideal}
\end{figure}

\cref{fig:train_friedmaan_ideal} and \cref{fig:train_boston_ideal} show the convergence results of different gradient estimation techniques for the Friedman and Boston housing datasets, respectively. The Guided-SPSA method outperforms the parameter-shift rule on the Friedman dataset and performs similarly to the parameter-shift rule on the Boston housing dataset. The Guided-SPSA outperforms standard SPSA-based gradient estimation in all scenarios. The same results are replicated on the validation and test metrics shown in \cref{tab:regression}. The convergence epochs shown in the \cref{tab:regression} are the average epoch at which the validation was lowest. It is to be noted that we did not use any early stopping method due to the significant variance in the convergence epoch on all the gradient estimation methods. Hence, all the methods were trained for a total of 100 epochs. SPSA-10, SPSA-20, and SPSA-30 mentioned in the \cref{tab:regression} denote the SPSA gradient estimation method with perturbation samples of 10, 20, and 30, respectively. SPSA with different $k$ were tested because the Guided-SPSA uses different perturbation samples at different stages of the training process. SPSA-max denotes that the perturbation sample $k$ matches the number of parameters in the VQC, resulting in the same computational complexity as the parameter-shift rule in terms of the number of circuit evaluations. The parameter-shift sample ratio $\tau$ of 0.5 and 0.7 was used for the experiments on the Friedman and Boston housing datasets, respectively. The SPSA damping constant $\epsilon$ was set to one for all the experiments with Guided-SPSA and ideal simulator as the convergence was stable without much instability. The results shown in \cref{tab:regression} are averaged over 25 training runs per method, and the variance of the results was in the order of $10^{-4}$.

\begin{figure}[htbp]
  \vspace*{-1em}\input{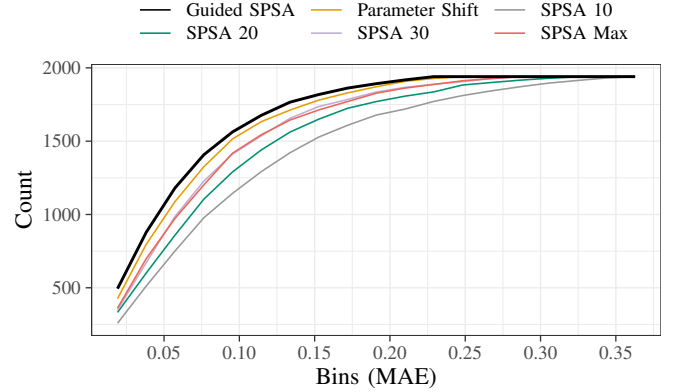}\vspace*{-2em}
     \caption{Cumulative distribution of the test samples from the Friedman dataset by different gradient estimation techniques.}
    \label{fig:test_friedman_cdf}
\end{figure}

Further, to get better insights into the performance of the models,  we investigated their outputs by examining the cumulative sum distribution of the MAE between the predicted and target values for the Friedman dataset.  In the results presented in \cref{fig:test_friedman_cdf}, we analyzed the cumulative sum distribution of the lowest MAE scores of 97\% of the data. We looked into the best 97\% data to filter out any outliers in each model. This was done to provide more informative results. Based on the findings in \cref{fig:test_friedman_cdf}, we concluded that the Guided-SPSA algorithm has the earliest saturation point. This indicates that, besides having the lowest average MAE, the Guided-SPSA also generalizes better than the other algorithms over the test dataset. The Guided-SPSA algorithm demonstrated similar or better performance than the standard parameter-shift rule while using 24\% and 18\% fewer circuit evaluations on the Friedman and Boston housing datasets, respectively.

\begin{figure*}[htbp]
    \vspace*{-1em}\input{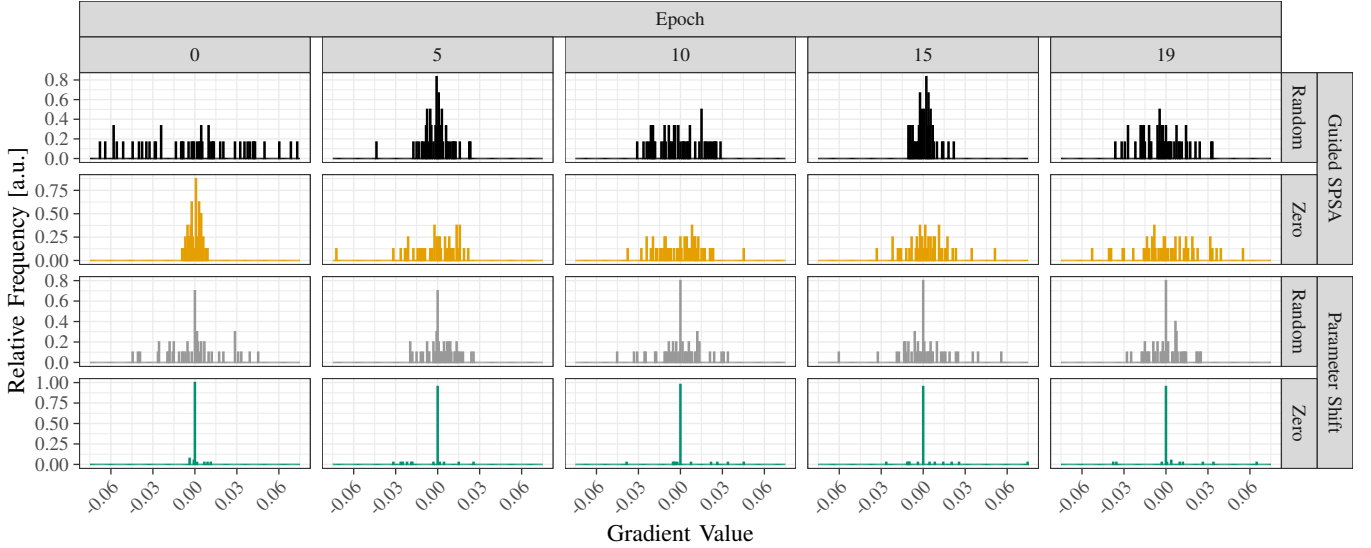}\vspace*{-2em}
    \caption{Comparison of gradient value distributions for G-SPSA and parameter shift
    approaches with different initialisation strategies (random initialization in range $(0, \pi)$, and zero initialization) for an increasing
    number of epochs.}\label{fig:histogram-grad}
\end{figure*}

\subsubsection{Shot-based Simulation}
\label{sec:shot-based}
For the next set of experiments, the same regression training runs explained in \cref{sec:ideal_simulation} were repeated using a shot-based simulator. Here, the given quantum circuit's expectation value is then approximated by simulating the circuit multiple times and building a probability distribution over the output of each simulation. The number of repetitions used to build the probability distribution is commonly known as "shots". We used 1024 shots to estimate each expectation value during these experiments. All the hyperparameters and the training procedure for these experiments were kept identical to the experiments in \cref{sec:ideal_simulation} except for the SPSA damping factor $\epsilon$. The $\epsilon$ was fixed to 0.5 instead of 1 as the shot-based simulator  introduces some inaccuracies in the estimation due to the finite number of shots used. Damping the magnitude of the gradients in the Guided-SPSA was necessary to facilitate stable training.  The results presented in \cref{tab:regression} demonstrate that Guided-SPSA performs similarly and has the same advantages when trained on a shot-based simulator as it does on an ideal simulator.

\subsubsection{Noisy Simulation}

The SPSA-based gradient estimation technique is often claimed to be well suited for noisy setups. In order to evaluate how well the algorithm would perform on real quantum hardware that is currently known to be noise-prone, we conducted a series of experiments on a noisy simulator that mimics the behavior of a quantum device affected by hardware noise. Here, we repeat the experiments described in \cref{sec:shot-based} with the same set of hyperparameters. To simulate a realistic NISQ device, we utilized the qubit connectivity and native gate set of the ``ibm\_brisbane"\cite{ibmq2022} hardware and incorporated the noise model supplied by qiskit for the same device. The validation and the test results presented in \cref{tab:regression} show that the Guided-SPSA algorithm exhibits similar convergence behavior and performance compared to ideal and shot-based simulation experiments.

\subsubsection{Noisy Simulation with Error-Mitigation}
\label{sec:error-mitig-sim}

 To study the convergence pattern and the performance of the Guided-SPSA algorithm under the impact of QEM, we conducted regression experiments with a noisy simulator, which employs ZNE. For this purpose, we used the native error mitigation capabilities of the IBM Quantum services and the cloud-based ``ibaq\_qasm\_simulator"\cite{ibmq2022}. Due to the high computational demand of ZNE and slow cloud-based simulators, we only studied the effects of the parameter-shift rule and Guided-SPSA based gradient estimation techniques under this setup. The results presented in \cref{tab:regression} indicate that Guided-SPSA exhibits similar convergence behavior to other experiments. It reaffirms the advantages in terms of optimal solution and fewer circuit evaluations.

\subsection{Suboptimal Initialization}

\begin{figure}[htbp]
  \vspace*{-1em}\input{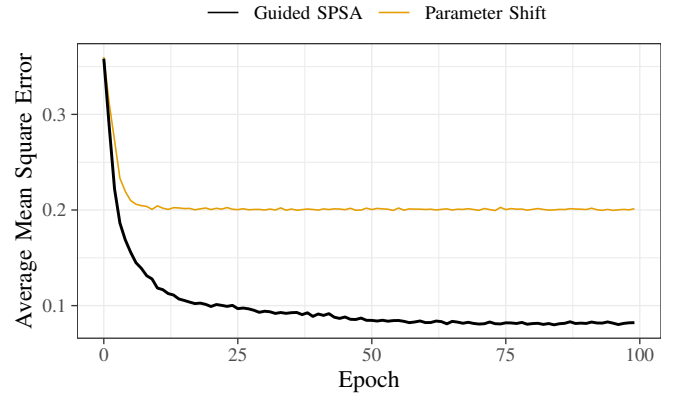}\vspace*{-2em}
  \caption{Convergence of parameter-shift rule and the
  Guided-SPSA based gradient estimation under sub-optimal
  all-zero initialization.}\label{fig:train_subopt}
\end{figure}

Eric~\etal~\cite{Anschuetz2022} showed that the VQCs struggle to reach the global minima with a bad guess of the initial parameter set. The term \enquote{bad} is, of course, relative and depends on multiple factors, such as the loss landscape of the VQC, dataset characteristics, hyper-parameters used for training, the algorithm itself, etc. One such bad guess for the Friedman dataset and the problem setup explained in \cref{sec:ideal_simulation} is zero initialization (\ie, all parameters are initialized to zero). Zero initialization is not always a bad guess, as Franz~\etal~\cite{franz2022uncovering} showed that it can help to achieve a faster convergence in the context of a reinforcement learning problem. However, for the regression task on the Friedman dataset using parameter-shift gradient estimation, the convergence leads to a suboptimal solution as shown in \cref{fig:train_subopt}. On the other hand, the same regression task with the Guided-SPSA gradient estimation converges to the optimal solution as in the random initialization case. To analyze this behavior, we studied the gradient distribution during early training stages for the Friedman dataset problem with the bad initial guess mentioned above: \cref{fig:histogram-grad} shows results for all combinations of random and zero initialization for Guided-SPSA and parameter shift rule. 

The distribution of gradient values remains concentrated sharply around zero throughout the early epochs when the VQC is zero initialized and the parameter-shift rule is used for gradient estimation. In contrast, using
zero initialization with Guided-SPSA delivers a gradient distribution that
follows the pattern of gradient distribution with random initialization
for both approaches.  We hypothesize that the randomness introduced by the SPSA part of the Guided-SPSA with a small perturbation sample size $k$ at the early stages of the training results in a wider gradient distribution and leads to a better convergence.

\subsection{Classification}

\begin{figure}[tbp]
  \vspace*{-1em}\input{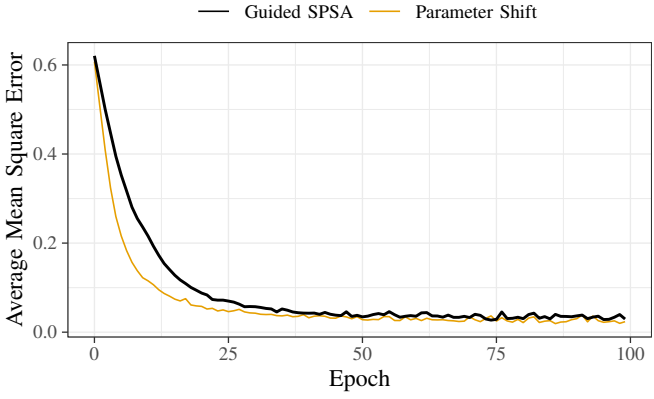}\vspace*{-2em}
  \caption{Convergence of different gradient estimation techniques using an ideal simulator for iris dataset classification.}\label{fig:train_classification}
\end{figure}

 To study the convergence behavior of the Guided-SPSA on classification problems, we trained a VQC to solve an according problem and compared its performance against the parameter-shift rule. As target, we chose the widely employed iris classification reference dataset. It consists of three different types of irises defined by four features and 
 contains 150 total samples. The VQC shown in \cref{fig:VQC_4qubit} is chosen to act as function approximator. There was no need for increment data updating layers in the architecture as the input features are four-dimensional, so it was omitted. The problem is a three-class classification problem; hence, the observables ZIII, IZII,  and IIZI were chosen to decode the output from the VQC, followed by a sigmoid activation function to predict the class probabilities. The results presented in \cref{fig:train_classification} show that the Guided-SPSA depicts similar performance to the parameter-shift rule in terms of classification. The same was observed in the test, as well as in the validation accuracy of the trained models. The gradient estimations through the parameter-shift rule utilized a total of $~$2.6 million circuit evaluations over the course of training, whereas the Guided-SPSA used $~$2 million circuit evaluations. A reduction of 25\% in the total number of circuit evaluations required to attain similar performance compared to the parameter-shift rule.

\subsection{Reinforcement Learning}

\begin{figure}[htbp!]
  \vspace*{-1em}\input{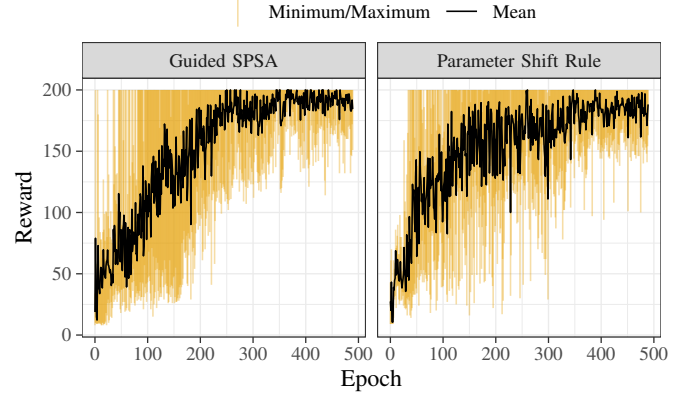}\vspace*{-2em}
  \caption{Training behaviour of a QRL agent for Guided SPSA and PS rule.}\label{fig:train_reinforcement}
\end{figure}

Finally, to get a complete picture of the different problem categories of the QML, we pit the Guided-SPSA algorithm against the parameter-shift rule in the quantum reinforcement learning (QRL) setting. We chose the task of solving the Cart-Pole-v0 environment \cite{Brockman2016} using the agent proposed by Franz~\etal \cite{franz2022uncovering} and the corresponding implementation provided in \cite{Franz_rl_repo} as the test bed for Guided-SPSA. As the algorithm proposed by Franz~\etal does not divide the training by epochs, the SPSA sample increment factor $\gamma$ from \cref{alg:guided-spsa} was calculated based on maximum steps of training instead of $N_{\text{epochs}}$. The results shown in the \cref{fig:train_reinforcement} show that the agents training using both the parameter-shift rule and the Guided-SPSA can learn an optimal policy and depict similar learning behavior. However, the parameter-shift rule used a total of 12.5M circuit evaluations during the course of training, whereas the Guided-SPSA used  9.6M circuits. A 24\% reduction compared to the parameter-shift rule.
\section{Conclusion}
We have introduced a novel gradient estimation scheme, Guided-SPSA, that combines advantages of the parameter-shift rule and simultaneous perturbation stochastic approximation based gradient estimation method. The algorithm features stable convergence, similar to or better than the parameter-shift rule, while incorporating benefits of SPSA such as fewer circuit evaluations. Guided-SPSA results in a 15\% to 25\% reduction in the number of circuit evaluations required throughout the course of training compared to the parameter-shift rule. It is
also well suited for NISQ devices, as it not only reduces the number of circuit evaluations required for good convergence, but also helps VQCs to avoid local minima and obtain more evenly distributed gradient values in certain problem scenarios. We
have empirically validated, using numerical experiments, the advantages of Guided-SPSA for regression, classification, and reinforcement learning tasks on widely employed reference
datasets and scenarios. The performance gain and the convergence pattern remained the same even when the model was simulated using a shot-based simulator, a noisy simulator, or under the presence of  error mitigation, at least for instance sizes that can be managed on current NISQ hardware. We anticipate these benefits will grow with problem size, and the Guided-SPSA gradient estimation approach will be a feasible and advantageous alternative to the parameter-shift rule or simultaneous perturbation stochastic approximation-based gradient estimation.

\section{Code availability}
The implementation of the regression and classification experiments are provided in the \href{https://github.com/maniraman-periyasamy/guided-spsa}{\textcolor{teal}{GitHub repository}}~\cite{guided_spsa_repo}. We have provided a DOI-safe reproduction package~\cite{mauerer:22:q-saner}: \href{https://doi.org/10.5281/zenodo.13137706}{\textcolor{teal}{10.5281/zenodo.13137706}}.
\newpage
\bibliographystyle{IEEEtran}
\bibliography{references}

\end{document}